\pdfoutput=1
\documentclass[fleqn,usenatbib]{mnras}
\usepackage{newtxtext,newtxmath}
\usepackage[T1]{fontenc}
\usepackage{ae,aecompl}
\usepackage{graphicx} 
\usepackage{amsmath} 
\usepackage{color}
\newcommand{\kms}{km\,s$^{-1}$}

\newcommand{\beii}{Be\,{\sc ii}}
\newcommand{\bevii}{$^{7}$Be}
\newcommand{\beviiii}{$^{7}$Be\,{\sc ii}}

\newcommand{\lii}{Li\,{\sc i}}

\newcommand{\livii}{$^{7}$Li}
\newcommand{\liviii}{$^{7}$Li\,{\sc i}}

\newcommand{\cai}{Ca\,{\sc i}}
\newcommand{\caii}{Ca\,{\sc ii}}

\newcommand{\nai}{Na\,{\sc i}}
\newcommand{\ki}{K\,{\sc i}}

\newcommand{\mgii}{Mg\,{\sc ii}}
\newcommand{\hi}{H\,{\sc i}}
\newcommand{\hei}{He\,{\sc i}}

\newcommand{\heii}{He\,{\sc ii}}

\newcommand{\oi}{O\,{\sc i}}
\newcommand{\oii}{O\,{\sc ii}}
\newcommand{\oiii}{O\,{\sc iii}}

\newcommand{\sii}{S\,{\sc ii}}
\newcommand{\feii}{Fe\,{\sc ii}}

\newcommand{\crii}{Cr\,{\sc ii}}

\begin{document}
\title[]{  \bevii\ detection in the 2021 outburst of RS Oph}
\author[]{Molaro, P. $^{1,2}$\thanks{E-mail: paolo.molaro@inaf.it }\thanks{Based on  data from Paranal Observatory, ESO, Chile}, Izzo, L.$^{3}$,  Selvelli, P.$^{1}$,
 Bonifacio, P.$^{5}$, Aydi, E $^{4}$, Cescutti, G. $^{1,2,10}$
\newauthor
Guido, E. $^{6}$, Harvey, E.J. $^{7}$, Hernanz, M. $^8$, Della Valle, M. $^9$
\\  
 $^{1}$  INAF-Osservatorio Astronomico di Trieste, Via G.B. Tiepolo 11, I-34143 Trieste, Italy\\
 $^{2}$  Institute of Fundamental Physics of the Universe, Via Beirut 2, Miramare,   Trieste, Italy\\
  $^{3}$ DARK, Niels Bohr Institute, University of Copenhagen, Jagtvej 128, 2200 Copenhagen, Denmark\\
  $^{4}$Center for Data Intensive and Time Domain Astronomy, Department of Physics and Astronomy, Michigan State University, East Lansing, MI 48824, USA\\
 $^{5}$ GEPI, Observatoire de Paris, Universit{\'e} PSL, CNRS, Place Jules Janssen, 92195 Meudon, France\\
 $^{6}$ Telescope Live, Spaceflux Ltd, 71-75 Shelton Street, Covent Garden, London, WC2H 9JQ, UK\\
  $^{7}$ Astrophysics Research Institute, Liverpool John Moores University, Liverpool, L3 5RF, UK \\
 $^8$ Institute of Space Sciences (ICE, CSIC) and IEEC, Campus UAB, Cam{\'i} de Can Magrans s/n, 08193 Cerdanyola del Valles (Barcelona), Spain\\
 $^9$ Capodimonte Astronomical Observatory, INAF-Napoli, Salita Moiariello 16, 80131-Napoli, Italy\\
 $^{10}$ INFN, Sezione di Trieste, Via A. Valerio 2, I-34127 Trieste, Italy\\
 $^{11}$Dipartimento di Fisica, Sezione di Astronomia, Università  di Trieste, Via G. B. Tiepolo 11, 34143 Trieste, Italy}
\date{Accepted.... Received 2022...}
\pagerange{\pageref{firstpage}--\pageref{lastpage}} \pubyear{2002}
\maketitle
\label{firstpage}
\begin{abstract}
The  recurrent nova    RS Oph underwent  a new outburst 
on  August 8,  2021, reaching a visible brightness of $V = 4.8$ mag. 
Observations of the 2021 outburst made
with the high resolution UVES spectrograph at the Kueyen-UT2 telescope of  ESO-VLT in Paranal  enabled  detection of   the possible presence of \bevii\ freshly made in the thermonuclear runaway reactions. The  \bevii\ yields can be estimated in   N(\bevii)/N(H) 
= 5.7 $\cdot 10^{-6}$,   which are     close to the  lowest yields measured 
in classical novae so far.
\bevii\ is short-lived  and decays only into \livii. 
 By means of a spectrum taken during the nebular phase   we estimated  an ejected   mass   of $\approx$ 1.1 $\cdot$ 10$^{-5}$  M$_{\odot}$, providing an  amount of $\approx$  4.4 $\cdot $ 10$^{-10}$ M$_{\odot}$ of \livii\   created in the 2021 event.    Recurrent novae of the kind of RS Oph   may synthesize  slightly  lower   amount of    \livii\   per event as   classical novae,  but  occur 10$^3$ times more frequently. The recurrent novae   fraction is in the range of 10-30 \%  and  they  could have  contributed to the  making   of   \livii\ we observe today.
The detection of  \bevii\ in RS Oph provides further support to the recent suggestion  that  novae are the most  effective      source of \livii\ in the Galaxy.

\end{abstract}
\begin{keywords}
{stars: individual: RS Oph; stars: novae
-- nucleosynthesis, abundances; Galaxy: evolution -- abundances}
\end{keywords}
\section{Introduction}

White dwarfs (WD) in close binary systems accreting H-rich matter from their companion star can explode as novae \citep{BodeEvansbook,dellavalle2020A&ARv..28....3D,chomiuk2021ARA&A..59..391C}. All nova explosions are recurrent because the explosion does not disrupt the white dwarf and accretion is reestablished after the explosion; the recurrence periods are in general very long.  However, there are about a dozen novae with more than one recorded outburst, meaning that their recurrence periods are shorter than 100 years. These are the so-called Recurrent Novae (RNe, \citealp{Schaefer2010}), whereas the other novae are named Classical Novae (CNe). 
It is known that the companion star of the white dwarf in classical novae is a main sequence star, and the binary system is a cataclysmic variable (CV), whereas  an evolved star, e.g. a red giant, is the companion of the white dwarf    in a subclass of RNe , i.e., T CrB, RS Oph,  V3890
Sgr and   U 745 Sco\citep{anupama2020JApA...41...43A,kato2012BASI...40..393K}. Recurrence periods as short as decades imply higher mass-transfer rates onto the white dwarf and higher white dwarf masses close to the Chandrasekhar mass limit in RNe than in CNe. Therefore, RNe are considered good scenarios of type Ia supernova explosions \citep{livio1992ApJ...389..695L,Schaefer2010, mikolajewska2017ApJ...847...99M}, and this has been suggested also for RS Oph \citep{hachisu2001ApJ...558..323H,hernanz2008NewAR..52..386H}.
However, there is evidence that not all RNe may end up their life as a type-Ia SN \citep{selvelli2008A&A...492..787S}.

The RN RS Oph comprises a white dwarf with  mass  of 1.2-1.4  M$_{\odot}$ close to the Chandrasekhar limit and a  K4-M0 red giant with a relatively small  mass of about 0.68-0.80  M$_{\odot}$ revolving    with a period of 453.6$\pm$ 0.3 d \citep{brandi2009A&A...497..815B,mikolajewska2017ApJ...847...99M}.   The outbursts result from a hydrogen thermonuclear runaway (TNR) on the white dwarf surface as a consequence of mass transfer from the red giant.  Five  historical outbursts have been recorded  over a century providing a    frequency of  once every 15-20 years. The   more recent ones which occurred   in 1985 and 2006  have been intensively  studied with observations ranging from X-ray to radio wavelengths, and covering all phases from quiescence to outburst \citep{2008ASPC..401.....E}.

The mechanism of the explosion is the same as in classical  novae (CN).  The H-rich accreted material on top of WD  grows  until it reaches conditions   at its bottom to ignite H under degenerate conditions, first through the p-p chains and later - when $T > 2x10^7$ K - through the CNO cycle. Nuclear burning proceeds fast and without control, since degeneracy prevents expansion of the envelope, thus leading to a TNR. Some beta-unstable nuclei produced by the CNO cycle are transported by convection to the outer envelope, where they decay, releasing energy that leads to the expansion and ejection of matter at  velocities  of several  thousands of  \kms\ with a simultaneous brightening  by several magnitudes \citep{Gallagher1978}.

\citealp{Arnould1975} and \citealp{Starrfield1978}  suggested that  in the thermonuclear  process, a mechanism similar to that one  proposed   by Cameron and Fowler  to explain the \livii\  rich giants  \citep{Cameron1955,CameronFowler1971} could take place.  The 
reaction $^3$He($\alpha$,$\gamma$)$^7$Be   leads to the formation of  $^7$Be which, if
 transported by convection to cooler zones  with a time-scale shorter than its electron capture time,    survives  from destruction.
This  suggestion    was quantitatively elaborated by  \cite{Hernanz1996,Jose1998} but  was thwarted by the non-detection of \livii\ in the outburst spectra of CN \citep{Friedjung1979}.  After decades of observational failures,  the possible presence of $^7$Li 670.8 nm resonance line    was  reported in nova V1369 Cen \citep{Izzo2015}, and  the parent nucleus $^7$Be   was   recognized  in several CN \citep{Tajitsu2015,Tajitsu2016,Molaro2016,Izzo2018,Selvelli2018,molaro2020MNRAS.492.4975M,Arai2021ApJ...916...44A,molaro2021arXiv211101469M}.  
  $^7$Be   is short-lived (53 d) and its presence in the  outburst spectra  implies that it has been   freshly created in the TNR processes of the nova event. The general non-detection of neutral $^7$Li  in CNe could be explained considering that \bevii\  decays with a  capture of an internal K-electron and therefore  ends up as    ionized lithium whose ground-state transitions are outside  the optical range  and are not observable \citep{Molaro2016}. 
 \bevii\ decays     into an excited  \livii\ state  that  de-excitates to the ground state  producing    high-energy photons at 478 keV  \citep{Clayton1981ApJ...244L..97C,Gomez1998}. Several  attempts to detect the 478 keV line with Gamma-ray satellites have been unsuccessful, but the limits derived are  consistent with the expected emission values \citep{Harris2001,jean2000MNRAS.319..350J,Siegert2018,siegert2021arXiv210400363S}.

The astrophysical origin  of Galactic  lithium  still represents an open question  \citep{Fields2011}.  The \livii\  abundance   today   is much  higher than  the  primordial  value. This   requires    the existence of one or several     sources  which are not yet identified. Spallation processes  in the interstellar medium is an established   source but its contribution cannot be higher   than 10\% \citep{Davids1970}. Stellar 
sources  such as  AGB  stars,  red giants and/or  supernovae have been suggested    to be actively producing lithium \citep{Romano2001}. The recent    yields measured in CNe imply  a  \livii\  over-production      by  up to  four  orders of magnitude  greater than  meteoritic   and therefore CNe alone could make up most of the Galactic   \livii\  \citep{Molaro2016,Cescutti2019,molaro2020MNRAS.492.4975M,molaro2021arXiv211101469M}.

\begin{table}
\begin{center}
\caption{RS Oph: basic data (to be completed). References:  1) a \citet{brandi2009A&A...497..815B}, 2. \citet{GAIADR3}, 3. \citet{2021AJ....161..147B}
\label{tab1}}
\begin{tabular}{llll}
\hline
Property & Value  &  & Ref\\
\hline
System    & WD+M2IIIpe   &   & 1 \\
Period    &        453.6 $\pm$ 0.4 d & & 1 \\
RA   & 17 50 13,20   & &  \\
 DEC & -06 42 28,5 && \\
Parallax & 0.416$\pm$ 0.023& & 2\\
Distance   & 2404$\pm$ 160 pc   &  & 2    \\
Geometric distance & 2402 pc  & 2276 -- 2524 pc & 3 \\
Photogeometric distance & 2441 pc & 2219 -- 2650 pc & 3 \\
G (EDR3) & $10.43 \pm 0.010$ &  &     \\
\hline
\end{tabular}
\footnotetext{Source: }
\end{center}
\end{table}

\begin{table}
\caption{Journal of   observations in August 2021. The RS Oph explosion  is taken in  2459434.50 MJD, or 2021 Aug 08.50 (±0.01) \citep{munari2021arXiv210901101M}. The values of the Cross  Disperser specify the wavelength of UVES  central setting: CD1=346.0 nm, CD3=580.0 nm, CD2=437.0 nm and CD4=760.0 nm.  MJD refers to the start of the exposure.
We also include the last observation of 29 March 2022 used for the nebular phase. }
\label{tab2}
\scriptsize
\begin{tabular}{lrrrrrrr}
\hline
\hline
\multicolumn{1}{c}{{Epoch}} & 
\multicolumn{1}{c}{{Inst}} &
\multicolumn{1}{c}{{Date}} & 
\multicolumn{1}{c}{{Grism}} & 
\multicolumn{1}{c}{{slit}} &
\multicolumn{1}{c}{exp} & 
\multicolumn{1}{c}{airmass} &
\multicolumn{1}{c}{DIMM}  \\
\multicolumn{1}{c}{} & 
\multicolumn{1}{c}{} & 
\multicolumn{1}{c}{MJD 2400000} & 
\multicolumn{1}{c}{} &
\multicolumn{1}{c}{}&
\multicolumn{1}{c}{sec}& 
\multicolumn{1}{c}{}&
\multicolumn{1}{c}{arcsec}\\
\hline
\hline
1&UVES&59436.0743 &CD3 &0.4&    10   &  1.059 &  0.60 \\
9-Aug&&59436.0743 &CD1 &0.4&60   &  1.059 &  0.60 \\
1.6 d &&59436.0750 &CD3&0.4 &10   &  1.059 &  0.55 \\
&&59436.0757 &CD3&0.4 	&10& 1.060 &  0.57 \\
&&59436.0763 &CD3&0.4 	&10   &  1.061 &  0.59 \\
&&59436.0789 &CD4&0.3 	&15   &  1.063 &  0.66 \\
&&59436.0789 &CD2&0.3 	&15   &  1.063 &  0.66 \\
2&UVES&59437.0211 &CD3 &0.6&	 5   &  1.069 &  0.50 \\
10-Aug&&59437.0211 &CD1&0.6 &	30   &  1.069 &  0.50 \\
2.5 d &&59437.0217 &CD3&0.6 &	 5   &  1.069 &  0.50 \\
&&59437.0287 &CD3 &0.6&	 2   &  1.061 &  0.45 \\
&&59437.0288 &CD1 &0.6&	15   &  1.061 &  0.45 \\
&&59437.0293 &CD3 &0.6&	 2   &  1.061 &  0.47 \\
&&59437.0339 &CD4 &0.6&	 2   &  1.057 &  0.46 \\
&&59437.0339 &CD2 &0.6&	 5   &  1.057 &  0.46 \\
3&UVES&59439.2151 &CD1 &0.4&    360   &  2.049 &  0.53 \\
12-Aug&&59439.2163 &CD3 &0.4&	60   &  2.075 &  0.56 \\
4.7 d & &59439.2176 &CD3 &0.4&	60   &  2.103 &  0.61 \\
&&59439.2221 &CD4 &0.3&	60   &  2.208 &  0.67 \\
&&59439.2221 &CD2 &0.3&	60   & 2.208 &  0.67 \\
4&HARPSN & 59439.8703 &HR & &300    & 1.242  &   \\
13-Aug &&  &     HR & & 300& 1.242&    \\
5.3 d &&  &     HR & &300  & 1.242&   \\
5&UVES&59440.2081 &CD3 &0.4&	60   &  1.965 &  1.23 \\
13-Aug&&59440.2081 &CD1 &0.4&    360   &  1.965 &  1.23 \\
5.7 &&59440.2094 &CD3 &0.4&	60   &  1.990 &  1.23 \\
&&59440.210 &CD3 &0.4&	60   &  2.014 &  1.23 \\
&&59440.2147 &CD4 &0.3&	60   &  2.100 &  1.82 \\
&&59440.2147 &CD2 &0.3&	60   &  2.100 &  1.82 \\
6&UVES&59441.2152 &CD3 &0.4&    120   &  2.173 &  2.28 \\
14-Aug&&59441.2152 &CD1 &0.4&    480   &  2.173 &  2.28 \\
6.7 d&&59441.2171 &CD3 & 0.4&   120   &  2.221 &  1.95 \\
&&59441.2191 &CD3 & 0.4&   120   &  2.272 &  2.25 \\
&&59441.2228 &CD4 &0.3&    120   &  2.372 &  2.25 \\
&&59441.2228 &CD2 &0.3&    120   &  2.372 &  2.25 \\
7&FIES & 	59441.9051 & HR & &  300   &  1.230 & \\
15-Aug && & HR& & 300    &  1.230  &   \\
7.4 d && & HR & &  300   &   1.230 &   \\
8&HARPSN & 59443.9280 & & & 300   & 1.282  &   \\
17-Aug &&  &     & & 300  & 1.282 &  \\
9.5 d &&  &    &  & 300  & 1.282 &  \\
9&UVES&59444.9656 &CD3 &0.6&    120   &  1.137 &  1.79 \\
18-Aug&&59444.9656 &CD1 &0.6&    480   &  1.137 &  1.79 \\
10.5 d &&59444.9675 &CD3 &0.6&    120   &  1.132 &  1.67 \\
&&59444.9695 &CD3 &0.6&    120   &  1.127 &  1.69 \\
&&59444.9735 &CD4 &0.3&    120   &  1.116 &  1.79 \\
&&59444.9735 &CD2 &0.3&    120   &  1.116 &  1.79 \\
10&FIES & 59445.8997 & HR &&  360    &  1.235 & \\
19-Aug && & HR & &360      & 1.235 &  \\
11.4 d && &HR && 360       & 1.235 &  \\
11&UVES&59447.0107 &CD3 & 0.6&   120   &  1.054 &  0.99 \\
20-Aug&&59447.0108 &CD1 &0.6&    480   &  1.054 &  0.99 \\
12.5 d &&59447.0146 &CD3 &0.6&    120   &  1.053 &  0.90 \\
&&59447.0188 &CD3 &	0.6&60   &  1.051 &  0.77 \\
&&59447.0205 &CD3 &0.6&	30   &  1.051 &  1.01 \\
&&59447.0205 &CD1 &0.6&	90  &  1.051 &  1.01 \\
&&59447.0214 &CD3 &0.6&	30   &  1.051 &  0.92 \\
&&59447.0223 &CD3 &0.6&	30   &  1.051 &  0.96 \\
&&59447.0273 &CD4 &0.3&    120   &  1.051 &  1.34 \\
&&59447.0274&CD2 &0.3&    120   &  1.051 &  1.34 \\
&&59447.0303&CD4 &0.3&	30   &  1.052 &  1.42 \\
&&59447.0303&CD2 &0.3&	30   &  1.052 &  1.42 \\
12&UVES&59457.0385 &CD3 & 0.6&   120   &  1.086 &  0.87 \\
31-Aug&&59457.0385&CD1 &0.6&    480   &  1.086 &  0.87 \\
20.5 d &&59457.0405 &CD3 &0.6&    120   &  1.089 &  1.13 \\
&& 59457.0424&CD4&0.6&120& 1.103&0.84\\
&& 59457.0471&CD2&0.6&120&1.103&0.84\\
 & UVES& 59667.2807&CD1&3.0& 240&    1.584 &   0.51\\
29 March& & 59667.2826 & CD3&3.0& 120 &1.584& 0.50\\
232.8&  &59667.2939 & CD4 &3.0&400 & 1.472 & 0.42\\
&  &59667.2940 & CD2 &3.0& 400 & 1.472 & 0.42\\
\hline

\end{tabular}
\end{table}

\begin{figure}%
\centering
\includegraphics[width=0.49\textwidth]{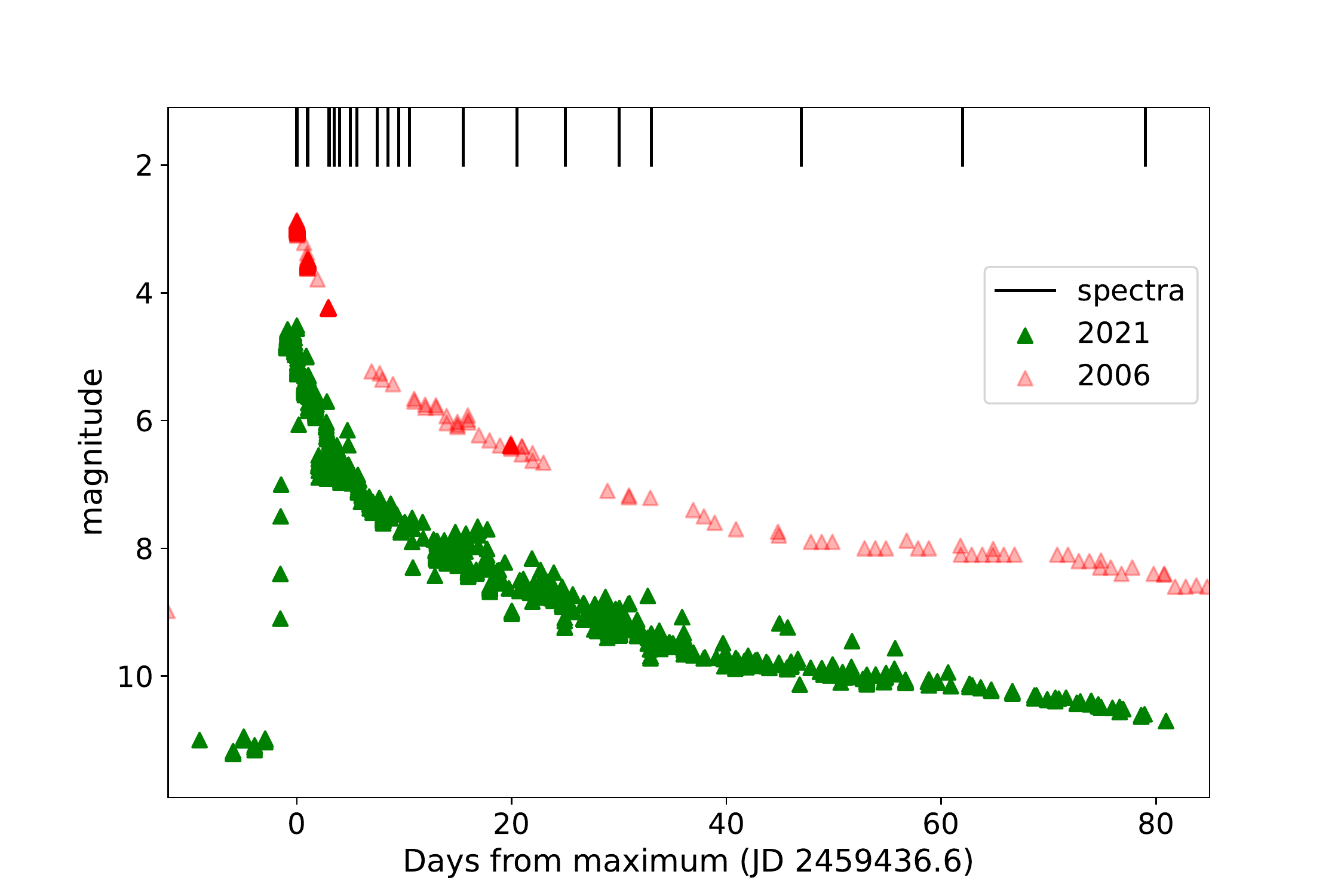}
\caption{ The V-band light curve of RS Oph as obtained by AAVSO members during the last outburst (green data) and the previous one in 2006 (red). Black lines mark the epochs of our spectral campaign.}\label{fig1}
\end{figure}

\begin{figure}
\centering
\includegraphics[width=0.5\textwidth,angle=0]{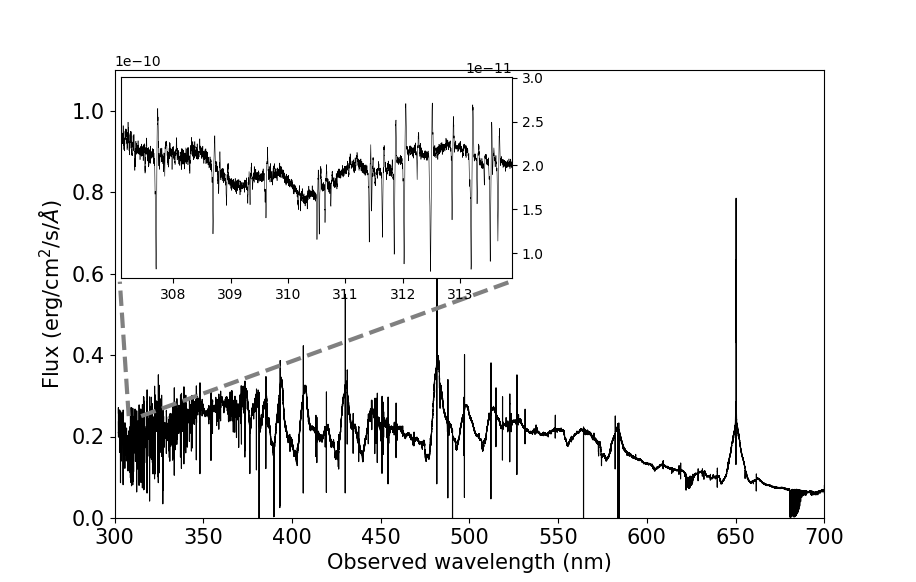}
\caption{Optical de-reddened spectrum 1.6 day after explosion (MJD 59434.47).  Broad P Cygni profiles of Balmer \hi, \hei, \feii, and \nai, dominate the spectrum. Narrow P Cygni   components are  superimposed to the broad emission lines of the nova outburst components originating in the  wind of the red giant. Towards the blue end they are so numerous that they mimic a noisy spectrum. The small portion in the \bevii\ regions is zoomed in the inset showing the  lines of the wind from the giant whose identifications are reported in Tab \ref{tab3}. }\label{fig2}
\end{figure}

\begin{figure}%
\centering
\includegraphics[width=0.4\textwidth]{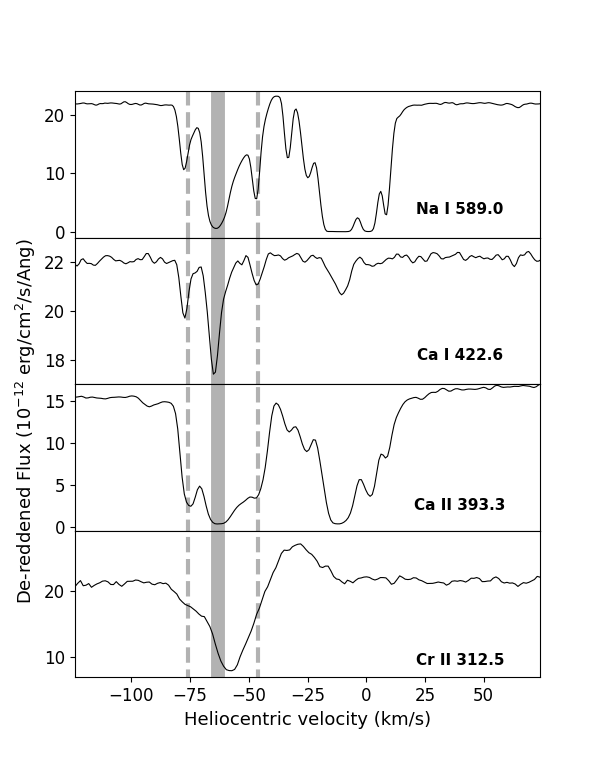}
\caption{ Spectra of day 1.5 of \nai\, 588.995 nm, \cai\, 422.6728 nm, and \caii\ K 393.3663 nm  showing the red-giant wind  and the interstellar medium structure. Velocities are heliocentric corrected.   The  systemic velocity for the RS Oph system is  -40.2 \kms \citep{fekel2000AJ....119.1375F}, and therefore the main component of the red giant wind is expanding with a velocity of about -23 \kms  in the RS Oph rest frame while  the two satellite ones at $\pm$  15 \kms with respect the main component       result from  the   orbital motion (dashed lines). The CrII 312.5 nm line  is  shown as example of the  lines  from  the circumstellar material produced by   the  red-giant wind  and  excited   by the nova  UV-flash.    }\label{fig3}
\end{figure}

\begin{figure}%
\centering
\includegraphics[width=0.5\textwidth]{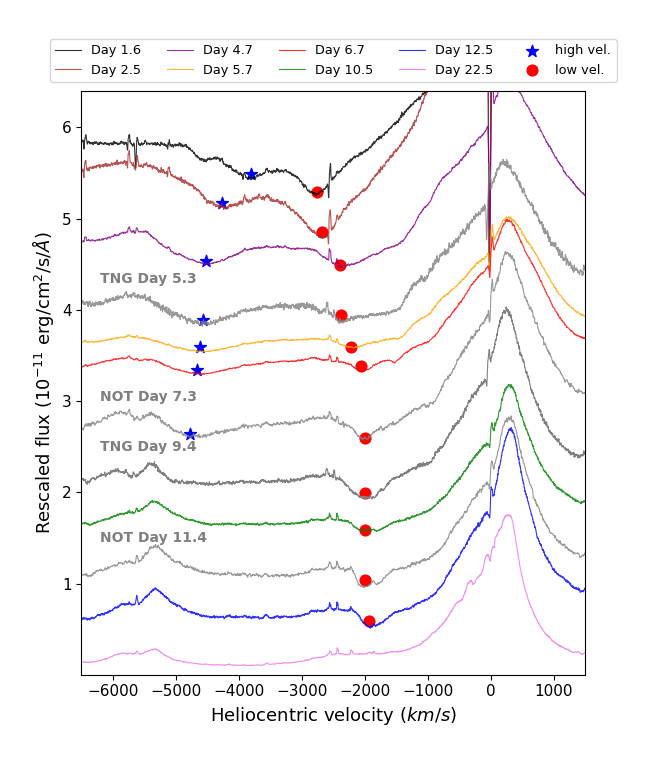}
\caption{The  absorption in H$_{\delta}$ 410.1735 nm in the outburst ejecta showing several components and their evolution, top to bottom,   in the first 23 days after explosion. The narrow P-Cygni feature with the absorption centered at at -63.4 \kms\ is the H$_{\delta}$ line of the red-giant wind. Note the \hei\ 402.6 nm emission line growing after day 10 at approximate position of -5300 \kms in the figure. The data show spectra taken with UVES at the VLT with different colors, while spectra obtained with HARPS-N at the TNG, and with FIES at the NOT are plotted in gray.   The high- and low-velocity components, as explained in the main text, are reported with blue stars and red circles, respectively.}
\label{fig4}
\end{figure}

\section{The 2021 outburst}\label{sec2}

The most recent RS Oph eruption occurred on 2021 Aug 08.50 (±0.01), or   2459434.50 MJD, reaching a  maximum V-band magnitude of 4.8  on  2021 Aug 09.58 (±0.05), or 2459435.68 MJD  \citep{munari2021arXiv210901101M}. Basic information on the RS Oph system is summarized in Tab \ref{tab1}. 
The parallax of RS Oph has been measured by Gaia 	
\citep{GAIADR3}: $0.373\pm 0.023$ mas. \citet{2021AJ....161..147B} have provided geometric and photogeometric distances
for almost 1.5 billions stars in the Gaia EDR3 catalogue. 
Both these values, as well as the lower and upper limits to these distances are provided in Table \ref{tab1}.
The Gaia parallax has to be corrected for the zero point \citep{2021A&A...649A...4L}, providing 0.416 mas.
A direct inversion of this parallax provides a distance of (2.40 $\pm$ 0.16) kpc.
The parallax measurement is  disturbed by the motion of the photocentre, due
to the orbital motion of the system as shown by the   astrometric excess noise 
that is 0.13 mas. A much better measurement of the parallax will be provided at the end of the Gaia mission,
where also the astrometric orbit shall be solved. Yet, Gaia EDR3 provides data that have been taken in almost 4 years
(25 July 2014 to 28 May 2017), thus averaging this motion over more than three orbits, implying that
the parallax should be accurate,  within the stated error. It is significant
that all three distance estimates (by parallax inversion, geometric with prior and  photogeometric with
prior) are consistent, within errors, with the distance derived from the expansion velocity
of the shells by \citet{Rupen_2008} that has been
suggested to be the best distance estimate of RS Oph by
\citet{magic2022arXiv220207681M}.

The 2021 outbursts   displayed  a rapid rise in brightness  reaching  about 5th magnitude within  24 hours from a  pre-burst  magnitude of 12.5 in the $B$ band\footnote{https://www.aavso.org/}.   The following photometric behaviour  was very similar  to  previous  outbursts as shown in Fig. \ref{fig1}. The behaviour  shows  fast decline  at the beginning and  slowing down  during  a second phase.  
The outburst  was  detected  all across the electromagnetic spectrum from radio  \citep{Sokolovsky2021ATel14535....1S,sokolovsky2021ATel14886....1S}  to X-rays \citep{enoto2021ATel14850....1E,enoto2021ATel14864....1E,ferrigno2021ATel14855....1F,luna2021ATel14872....1L,page2021ATel14848....1P,page2021ATel14894....1P,rout2021ATel14882....1R,shidatsu2021ATel14846....1S}, gamma-rays   \citep{cheung2021ATel14834....1C,cheung2021ATel14845....1C, magic2022arXiv220207681M} at Gev energies and, for the first time, even at TeV energies \citep{wagner2021ATel14844....1W,wagner2021ATel14857....1W}. Search for neutrino emission with IceCube was negative \citep{pizzuto2021ATel14851....1P}.
The  radio emission is largely nonthermal \citep{sokolovsky2021ATel14886....1S}. An inverted spectrum shape that was observed early in the eruption was produced by  external free-free absorption or  synchrotron self-absorption within the radio emitting region. A nearly flat spectrum  together with deviations from a simple power law fit are observed at later times indicating that  the emitting region is inhomogeneous or remains partly hidden behind some absorbing material.

The spectra of RS Oph were obtained at VLT/UVES by triggering an ESO ToO program (Prog. ID: 105.D-0188, PI P. Molaro), after the  alert  with the earliest optical spectrum taken at MJD 59436.07427, or 1.6 days after explosion.  The  settings used were DIC1 346-564,  with ranges 305-388\,nm    and  460-665\,nm,  and    DIC2 437-760 with ranges 360-480\,nm    and  600-800\,nm  in the blue and red arms, respectively.    The journal of the observations for the  nova is  provided in  Table \ref{tab2}. 
The nominal resolving power  of early spectra was of  $R= \lambda /\delta \lambda  \approx 100,000$ for the blue arm   setting the slit at 0.4 arcsec and $\approx 130,000$ for the red arm for a slit of 0.3 arcsec. In late spectra the   slit was set to 0.6 arcsec    for the blue arm to cope with the nova fading in the blue providing a uniform    $R= \lambda /\delta \lambda  \approx 66,000$ in both arms. The actual slits used are  provided in  Table \ref{tab2}.  Overlapping spectra were combined for each epoch to maximise the signal-to-noise ratio. The spectra have been carefully cleaned from the telluric O$_3$ Huggins bands by means of a B-type subdwarf HD 149382 \citep{schachter1991PASP..103..457S}.
Few  spectra were also   obtained with HARPS-N,   The High Accuracy Radial velocity Planet Searcher for the Norther hemisphere \citep{cosentino2012SPIE.8446E..1VC} (Prog. ID: A43-TAC20 and A44-TAC17, PI: Izzo) at the 3.6m Telescopio Nazionale Galileo (TNG) and with FIES, the high resolution FIber-fed Echelle Spectrograph \citep{telting2014AN....335...41T}, at the 2.5 m Nordic Optical Telescope (NOT), both at la Palma, Spain (Prog. ID: 63-013, PI: Izzo). The spectral range of the HARPS-N is from 383 nm to 693 nm with a  $R= \lambda /\delta \lambda  \approx 120,000$. For FIES the spectral coverage is 400-830 nm  with a  $R= \lambda /\delta \lambda  \approx   65,000$. 
  The spectra have been flux calibrated by means of  spectroscopic standards and corrected for a   reddening  of $E(B-V) = 0.73$ \citep{cassatella1985ESASP.236..281C,snijders1987Ap&SS.130..243S} The flux calibration for the late,  nebular spectra has been refined using photometry from AAVSO\citep{Kafka2021}.

A portion of the first optical spectrum  from 300 to 700 nm is shown in  Fig. \ref{fig2}.  Broad P Cygni profiles of Balmer \hi, \feii, \oi, and \nai, along with weaker, broad lines of \hei\ and  dominate the spectrum.  The \feii\ multiplet 42  is the main contributor to the emission features at 492.4, 501.8,  and  516.9 nm.   in Fig \ref{fig4} the  several absorption components and their evolution  in the H$_{\delta}$ 410.1735 nm are shown.  The blue edge of hydrogen absorption  shows a  maximum terminal  velocity of the ejecta    of $\sim$ -4700 \kms\ on day 1.6, which then accelerates to  $\sim$ -5200 \kms\  at day 2.5 and  remains constant afterwards.  This is a quite different behavior from what described in \citet{magic2022arXiv220207681M}   where   the terminal velocity is taken to reach -4200 $\pm 250$ \kms   with a notable decrease to about -2000 \kms   around day 4. As it is possible to see in Fig \ref{fig4} the component at about -2000 \kms was present since the beginning of the outburst and does not result from a decrease of the high velocity component. 
 \citet{tatischeff2007ApJ...663L.101T} from the analysis of the X-ray and IR observations of the RS Oph 2006 eruption predicted that a recurrent nova  with a red giant companion  can indeed accelerate protons and electrons, as now has been suggested by high-energy detection \citep{magic2022arXiv220207681M,2022arXiv220208201H}.  
 
The broad emission lines  are  shrinking      and   H$_{\alpha}$  becomes   highly non Gaussian    revealing  a  signature of  bipolar flow.  An expanding bipolar  structure was   detected after the 2006 eruption in the E-W direction and a similar circumstance could occur also for that of  2021 \citep{ribeiro2009ApJ...703.1955R, montez2021arXiv211004315M}.  A detailed description of spectral evolution will be given elsewhere.

Narrow P-Cygni   components which originate in the  wind of the red giant  are  superimposed to the broad emission lines of the nova outburst. In Fig. \ref{fig3} are shown the  spectral regions  of \nai\,, \cai\,, \caii\ K, and \liviii\,,   showing the red-giant wind  and the interstellar medium structure. 
The \nai\ lines show 
 red-giant  wind   absorption components at ~-48, ~-63, and ~-77 \kms.   The  stronger  central component  at - 63 \kms\ and the two satellite ones        result from  the  systemic orbital motion which has an amplitude of about 30 \kms.  The  systemic velocity for the RS Oph system is  -40.2 \kms \citep{fekel2000AJ....119.1375F}, therefore the wind is expanding with a velocity of about -23 \kms  in the RS Oph rest frame.  In fact,   the  wind  velocities of the red giant wind measured in the recent outburst are very close to those measured in the 2006 outburst and in quiescence  \citep{patat2011A&A...530A..63P}.   A few lines falling in the   \beii\  and \nai\  regions  together with their velocities are reported in Tab. \ref{tab3}. These  lines form in  the circumstellar material produced by   the  red-giant wind  and are excited   by the nova  UV-flash. They  only  show evidence  of the  main  central component seen as a P-Cygni profile. It is plausible that the slow  red-giant wind  filled up the  cavity created by the previous  explosions and in particular by  that  of 2006.
˙
\begin{figure}%
\centering
\includegraphics[width=0.5\textwidth]{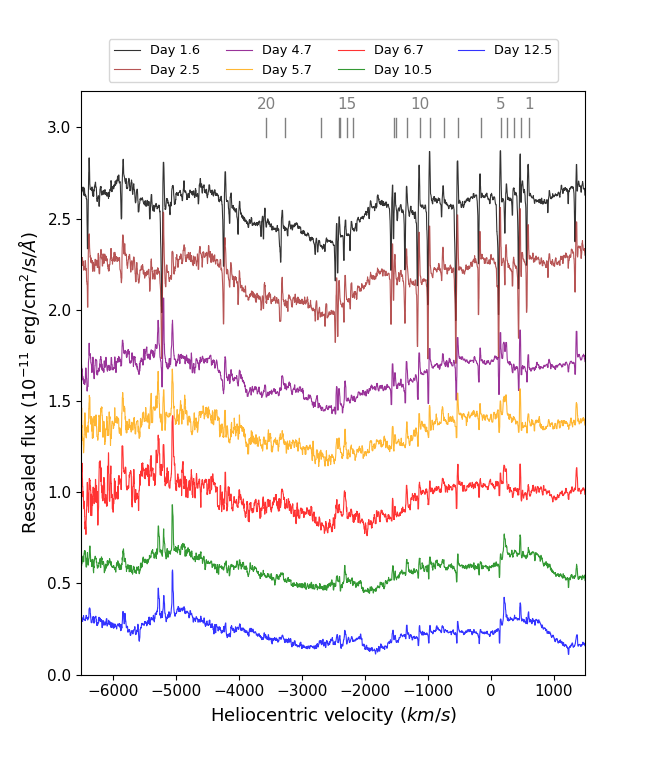}
\caption{As in the previous figure but for the \bevii\  region. The zero in the x-axis is taken at 313.0583 nm.  Spectra are corrected for an extinction of E(B-V)=0.73 and for  O$_3$ absorption. The spectra show the complex absorption due to the outburst ejecta and the narrow P-Cygni lines of the red-giant wind which are identified in Tab. \ref{tab3}. The  \oiii\ 313.37 nm begins to be seen in emission in the last spectra at slightly positive velocities.}
\label{fig5}
\end{figure}

\begin{figure}%
\centering
\includegraphics[width=0.5\textwidth]{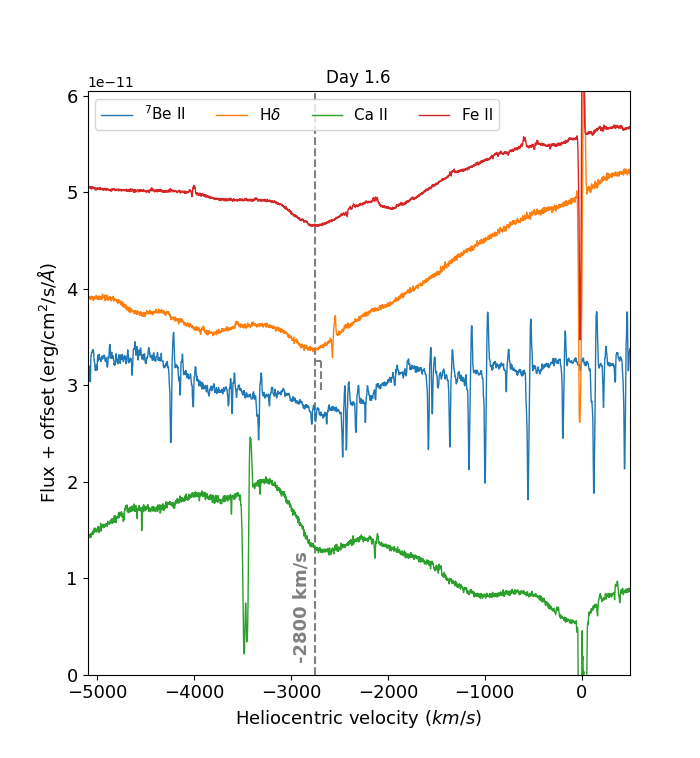}
\caption{ Spectra of \bevii\ together  with H$_{\delta}$, \caii\ K and \feii\  516.903   at day  1.6   plotted on a common velocity scale. For \bevii\ the zero of the scale is   at 313.0583 nm.  The emission falling in the middle of the   \caii\ K spectrum is    \hi\  388.9.05  nm   and  the two absorptions  at $\approx$  -100  and  $\approx$ -1100 \kms\ are the   components at -2800 and -3800 \kms\ of  \caii\ H  at  and  \hi $_{\epsilon} $ lines, 
respectively.}\label{fig6}
\end{figure}

\begin{figure}%
\centering
\includegraphics[width=0.5\textwidth]{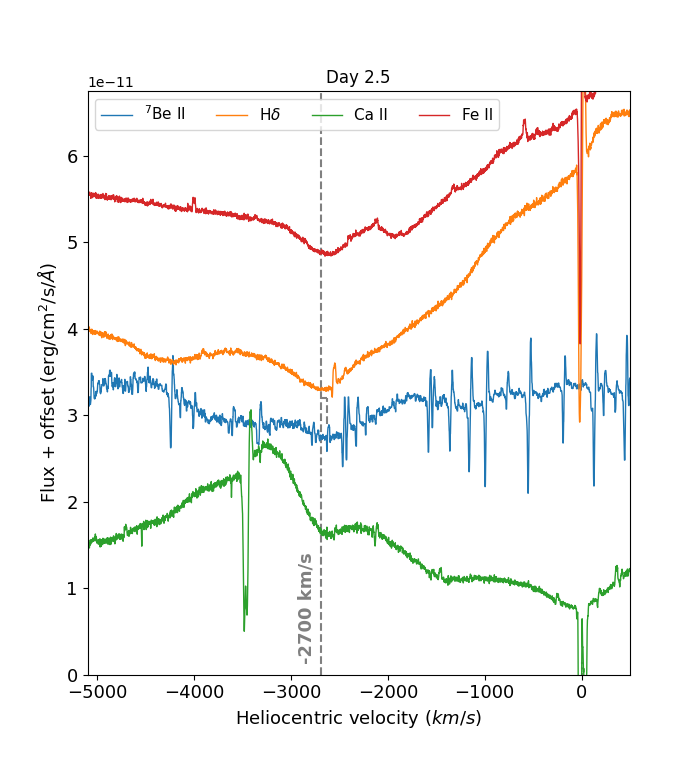}
\caption{ As in the previous figure for   day 2.5. To note the enhancement of the emission of  the  \hi\  388.905  nm  falling in the middle of the   \caii\ K absorption.  }\label{fig7}
\end{figure}

\begin{figure}%
\centering
\includegraphics[width=0.5\textwidth]{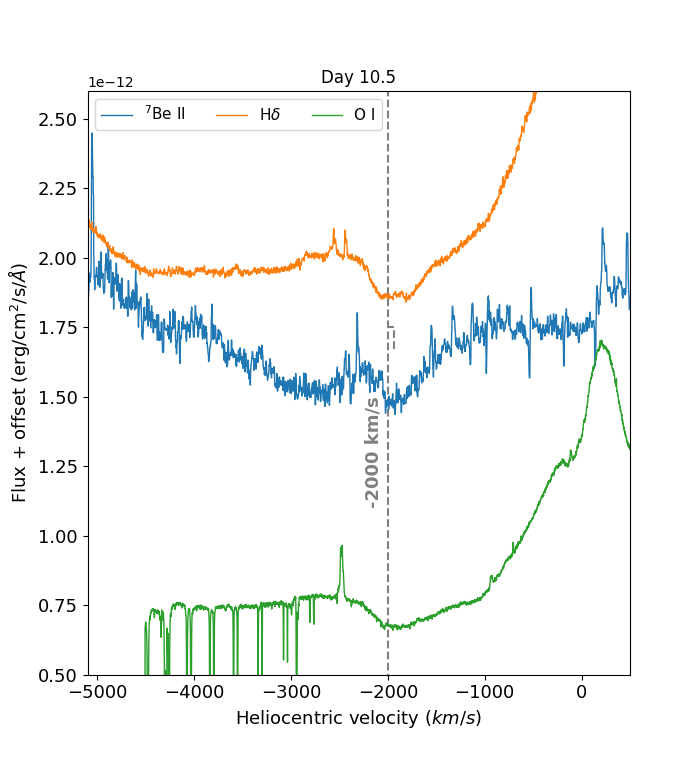}
\caption{ \bevii,  H$_{\delta}$ and   \oi\ triplet at 777.19, 777.41 and 777.54  nm  spectra at day 10.5  plotted on a common velocity scale. The \oi\ triplet is not resolved though with a separation of 133 \kms and the \oi\ 777.5388 nm  wavelength    is used for the zero velocity in the plot.  Note that the absorption of OI is affected at its left by the emission line of \feii\ (73) at 771.2 nm.    No other metal elements are detected  to check the common absorption system. The narrow absorption features present in the \oi\ spectrum are the telluric  lines of the molecular O$_2$  A band.}
\label{fig8}
\end{figure}

\begin{figure}%
\centering
\includegraphics[width=0.5\textwidth]{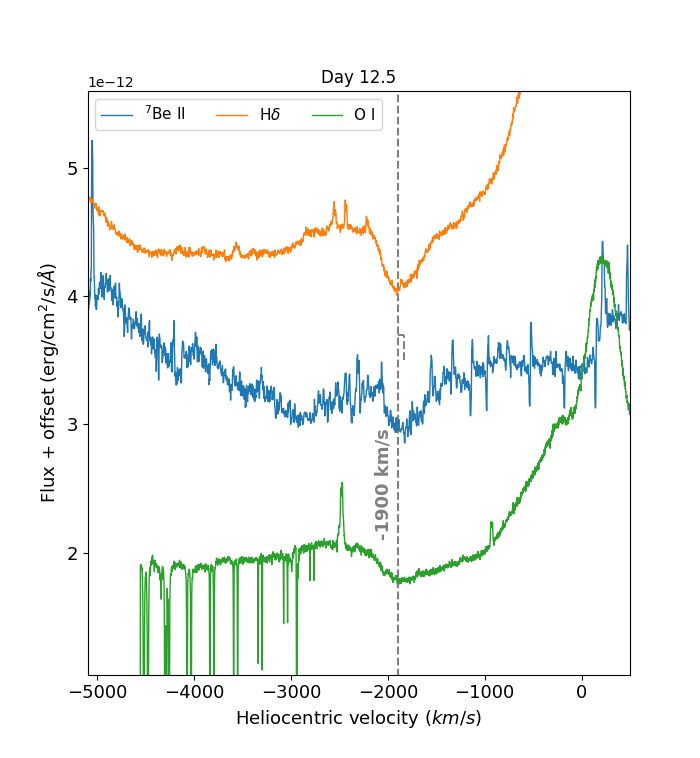}
\caption{ As in the previous figure. After  this epoch no more absorption is  detectable. }\label{fig9}
\end{figure}

 \begin{table}
\label{tab3}
\caption{Lines of the red-giant  wind in the \bevii\, \nai\, \feii\ 430.3 nm  and \hi\ $_\delta$ spectral regions as measured in the spectrum at day 1.6.  Some DIBs velocities are also given to emphasize the velocity structure of the interstellar medium towards RS Oph. 
Velocities are  heliocentric  with a correction value of -21.86 \kms. Last column reports the identification number of the lines shown in Fig.\ref{fig3}}
\scriptsize
\begin{tabular}{lrrrrrrr}
\hline
\hline
\multicolumn{1}{c}{{ident}} &
\multicolumn{1}{c}{{abs}} & 
\multicolumn{1}{c}{{fwhm}} & 
\multicolumn{1}{c}{vel} & 
\multicolumn{1}{c}{em} &
\multicolumn{1}{c}{fwhm}&
\multicolumn{1}{c}{ID }\\
\hline
\hline
\crii\	3136.686	            & 3136.327 & 0.17 &	-56.2& 3136.663 & 0.43  & 1\\
  \feii\  3135.362    & 3134.992 & 0.18 & -57.3 &  3135.329&  0.51 & 2\\
 \crii\	3134.303			            & 3133.963 & 0.11 &-54.4	& 3134.235 & 0.82  & 3\\
 \feii\  3133.048    		        & 3132.715 & 0.13 &-53.8	& 3132.999 & 0.25  & 4\\
  \crii\  3132.053    & 3131.662 & 0.23 & -59.3&  3132.022&  0.79 & 5\\ 
  \crii\  3128.700    & 3128.350 & 0.17 & -58.9&  3128.623&  0.21 & 6\\
  \crii\  3124.973    & 3124.586 & 0.23 & -59.0&  3125.004&  0.61 & 7\\
  \crii\  3122.602     & 3122.259 & 0.11 &-54.7	& 3122.532 & 0.25  & 8\\
  \crii\  3120.369   & 3119.989 & 0.18&-58.4	& 3120.334 & 0.89 & 9\\
  \crii\ 3118.649      & 3118.274 & 0.18 &-58.0	& 3118.621 & 0.97  & 10\\
  \feii\ 3116.580      & 3116.221 & 0.17 &-56.5 	& 3116.563 & 0.44  & 11\\
\feii\ 3114.683		            & 3114.359 & 0.13 &-53.1	&  3114.620&  0.22  & 12\\
  \feii\ 3114.295     & 3113.936 & 0.16 &-56.5	&  3114.177&  0.19  & 13\\
ScII 3107.52 		            & 3107.235 & 0.09 &	-49.3&  3107.550&  0.59  & 14\\
  \feii\ 3106.565     & 3106.229 & 0.13 &-54.3&  3106.505&  0.19  & 15\\
  \feii\  3105.554    & 3105.232 & 0.14 &-53.0	&  3105.502&  0.46 & 16\\
  \feii\  3105.168    & 3104.813 & 0.22 &-56.2	&  3105.046&  0.13 & 17\\
VII 3102.29		            & 3101.971 & 0.12 &	-52.7&  	   &	      & 18\\
KI  3102.05?	            & 3101.559 & 0.12&-69.4	&  	   &	      & 18\\
  \feii\   3096.294    & 3095.937 & 0.15 &-56.5	&  3096.295&  0.42 & 19\\ 
		            & 3093.161 & 0.09 &	&  3093.402&  0.22 & 20\\
		          \crii\ 3093.17?  & 3092.773 & 0.11&-60.4	&  3092.996&  0.23 & 20\\
		            & 	       & 	     &	&	   &	  &    \\
  \nai\   5889.95095  & 5889.451 & 0.07 &  -47.3&	   &	 &     \\
                    & 5889.151 & 0.46&  -62.6&	   &	 &     \\
		            & 5888.868 & 0.11&  -76.9&	   &	 &      \\
                    & 5889.727 & 0.07 &	-33.2&	   &	 &      \\
		            & 5889.903 & 0.14 &	-24.3&	   &	 &      \\
		            & 5890.167 & 0.73 &  -10.9&	   &	 &      \\
		            & 5890.384 & 0.35 &	+0.2&	   &	 &      \\
		            & 5890.533 & 0.16 &	+7.8&	   &     &      \\
\hi\ $_\delta$ 4340.472		            		            & 4339.87 & 0.43 &	-63.5&4340.557	   &0.38      &     \\
\feii\ 4303.168& 4302.672& 0.15&-56.5& 4303.040&0.50&\\		   
DIB 6660.71&6660.870 &0.63&-14.7 &&\\
DIB 6613.62 & 6613.865& 0.91& -10.8&& \\
DIB 5849.81&5849.914&0.63&-16.5&&\\
DIB 5797.060&5797.268&0.76& -11.1&&\\

\hline
\hline
\end{tabular}
\end{table}

\section{  \bevii\ detection \& abundance}

The   UVES, HARPS-N and FIES spectra covering  the H$_{\delta}$      are shown in Fig. \ref{fig4} for the first 11  epochs of our observations. Besides  emissions, the outburst hydrogen spectra  show  broad absorption  in a range of radial velocities spanning from -1000 to -5000 \kms. There are   two broad  absorption features with central velocities at  about -2800 \kms and -3800 \kms that show some  variation in  radial velocities from day 1.5 to day 12.5. To note the \hei\ 402.6 nm  emission at about -5200 \kms\ which is growing up in the last four spectra of the sequence. The next spectrum is  on the 30 August, or day 22.5 after explosion, where  all the absorption components  present in the Balmer lines are no longer visible, either because they weakened or because they are hidden in the plethora of emission lines developed by the nova.
The corresponding  UVES spectra centered onto the \bevii\ 313.0 nm  region are shown in Fig. \ref{fig5}  revealing  a quite complex behaviour. The narrow P-Cygni lines of the red-giant wind falling in this restricted spectral portion are identified and measured in Tab. \ref{tab3}. To note that while the \crii\ lines maintain a  P-Cygni profile   the \feii\  from day 5 onwards   are seen only in emission. 
This is because   the \crii\ are formed from transitions which start from metastable levels at about 2.46 eV,  while those of Fe II  originate from
 non-metastable levels with  an higher energy of about 3.9 eV. Thus, 
when radiation drops they   are depopulated earlier with a corresponding reduction or suppression of the absorption. To note  that the OIII line at 313.37 nm  appears in emission in from Day 12.5, as shown in Fig \ref{fig5}.

Fig. \ref{fig6} shows the \bevii~spectra  of day 1.5   along with the portions of  \caii\ K,  Fe II and H$\delta$  lines  on a common velocity scale.    H$\delta$  is adopted as representative of the absorption  seen in hydrogen being the less blended and still relatively strong.  \caii\ H  cannot be used since it is contaminated with H$_\epsilon$  and also  with the Galactic interstellar components of  \caii\ K.  To note that the hydrogen line \hi\  388.905  is falling in the middle of the   \caii\ K absorption and shows a P-Cygni profile with  narrow absorption and  narrow emission.      The   \hei\ 388.8605, 388.8646, 388.8649 nm lines are also present with the 
probable  emission  suppressed by  the \hi\   absorption falling on
the same position. A common broad absorption  with central velocity  of about -2800 \kms is present in all  species, and possibly also one  at   $\approx$ -3800 \kms. The vertical lines in the figure   mark  the  \beviiii~ doublet at $\lambda313.0583 + \lambda313.1228$  doublet  which has a  separation of 62 \kms.
To note that the -3800 \kms\  is absent in the very early spectrum  at  MJD 59435.46, or 0.96 days after explosion, of  \citet{taguchi2021}. The faster component at -3800 \kms\ appears only over than half a day later. This is something common in classical novae, where we see slow, pre-maximum P Cygni profiles, followed by faster components that  are delayed by a few hours to a few days  \citep{Aydi2020ApJ...905...62A}. \citet{Aydi2020ApJ...905...62A} interpreted low velocity components  of typically few hundreds \kms, originating in a slow, early ejection, and the fast component is a faster wind driven by residual nuclear burning and  expanding at velocities of thousands of \kms .   
Unfortunately, in the RS Oph  spectra there is no evidence of narrow components  in the outburst absorptions at any available epoch, which would  reveal the \bevii\ doublet structure  clearly demonstrating  the presence of \bevii. Thus, the identification with \beviiii\ 313.0 nm relies only on the similarity of line  profile and on the absence of alternative significant blends as it  has been discussed in other studies \citep{Tajitsu2015,Molaro2016}. Fig. \ref{fig7}  shows the same transitions  at day 2.5. The hydrogen line \hi\  388.9.05 nm develops a strong emission which modifies significantly  the \caii\ K absorption profile. Close  correspondence persists    for the -2800 \kms  velocity component which is  marginally  affected by the emission. By day 4.7 the absorptions of metal components are gone. 
Interestingly,  the   features attributed to   \bevii\ remain there, which implies  that the absorptions  cannot be made of a blend of metallic lines. At this epoch the highest velocity component in hydrogen lines showed  an acceleration, but the   correspondence with the analog \bevii\ absorption is lost. This could be the result of the shock wave following the interaction of  the expanding ejecta with the circumstellar material created by the wind of the red-giant and the subsequent proton acceleration \citep{magic2022arXiv220207681M}. The velocity coherence with hydrogen  is recovered when  a   component  shows up at   $\approx$  -1900 \kms  in the spectra after  days 6.7.  This  component is possibly the -2800 \kms decelerated by  the collision between high velocity gas of the outburst and the circumstellar material created by the red giant wind since the 2006 explosion.  H$\delta$ and  \bevii\ of the    last two epochs  are shown in Fig \ref{fig8} and \ref{fig9} together with \oi\ which is the only element strong enough to be seen in   absorption. 
The profiles of the new component are very similar and are  suggestive of the presence of a common absorption at these velocities.

 The resonance  \caii\ K  line is required to estimate the \bevii\ abundance under the assumption that the two ions   are  in their main ionization  stage. The possibility of overionization of \caii\ with respect to \beviiii\ has been discussed  in the case  of CNe and found  unlikely \citep{molaro2021arXiv211101469M}.   The absorption of the \caii\ K line could be seen  clearly only at days 1.6 and 2.5, but only  in the  spectrum taken at day 1.6 is not affected by the nearby emission,  so that the \bevii\ abundance could  necessarily  be derived   only at the first epoch.

\begin{figure} %
\centering
\includegraphics[width=0.5\textwidth]{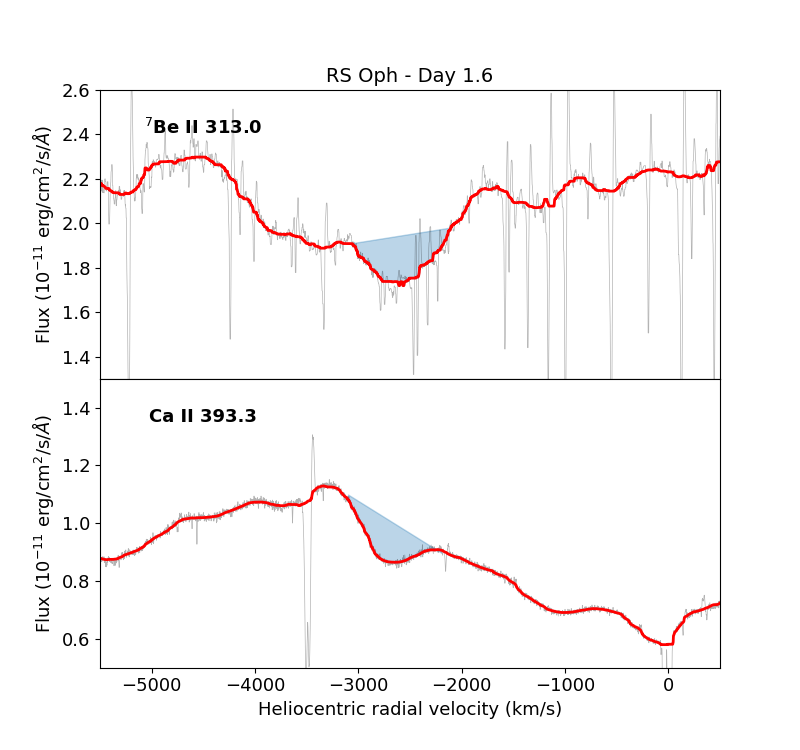}
\caption{   The \beviiii\ and the \caii\ K  spectra  taken at 1.6 d. The red lines show the spectra cleaned from the features of the circumstellar material made by the red giant wind. Highlighted are the conservative measurements of the EWs made by taking the local continuum as low  as possible. }\label{fig10}
\end{figure}

 The equivalent width (EW) of the sum of the \beviiii\,  lines is   compared with the \caii~  K line at 393.366 nm. Following  previous analysis we have  

\begin{eqnarray}\label{eq:1}
  \frac{N(\mbox{\beviiii})}{N(\mbox{\caii})}
  & = &   2.164   \times \frac{EW(\mbox{\beviiii\,, Doublet})}{EW(\mbox{\caii\,, K})} 
\end{eqnarray}

\noindent
 with $log (gf)$  of -0.178 and -0.479 for the \beviiii\ doublet, and    +0.135 for the \caii\,K  line  \citep{Tajitsu2015,Molaro2016}.
The measure of the equivalent width  of the  main  \beviiii\  absorption at -2800 \kms\     is performed  in a very conservative way in the spectrum of first epoch  as  shown in Fig. \ref{fig10} by the shadowed area. The EW  of the \beviiii\ doublet  is   of   1065 $\pm$ 86 m\AA.   The main  uncertainty  comes from  the continuum placement which is traced here as low as possible. Close to  \beviiii~ doublet there are several  lines mainly of \crii\ and \feii\  which could contribute to  the observed absorption. However,   the close correspondence of the 
absorption profile assures they are not the dominant species.  The equivalent width  of the \caii\ 393.3 nm  line also  illustrated in Fig.  \ref{fig10} is of  859 $\pm$ 116 m\AA,  providing   a  ratio of EW({\beviiii\,, Doublet})/EW(\mbox{\caii\,, K})  $\approx$ 1.20$\pm 0.2$.  The high velocity component is contaminated by the emission of the hydrogen line and likely the \caii\ K component is  partially filled up by the hydrogen emission. The values of the  higher velocity absorption would be of 947 m\AA, and 164 m\AA, respectively, providing a  ratio of   EW(\beviiii)/EW(\caii) $\approx$  4.5. We thus  consider the ratio from the lower velocity component less affected by blends and     more   representative of the  ratio in all components of the outburst material.

Using  Eq. \ref{eq:1},  we obtain   N(\beviiii)/N(\caii) $\approx$  2.61 $\pm 0.43$. Since  the  half life time decay of \bevii\  is of 53.22  days, after 1.6 days    from explosion, the    abundance   had  not significantly reduced assuming the TNR occurred  together with the explosion.  The   abundance  of  calcium is  taken solar N(Ca)/N(H) = 2.19 $ \pm 0.30 \cdot 10^{-6}$  \citep{lodders2019arXiv191200844L} which gives    an abundance of   N(\bevii)/N(H) 
$\approx$  5.7 $(\pm 1.5)  \cdot 10^{-6}$, or $X(\mbox{\bevii})/X(\mbox{H})$  
$\approx$  4.0 $(\pm 1.1)\cdot 10^{-5}$  in mass. Should Calcium abundance   be different from  solar,   the final N(\bevii) /N(H) would change  accordingly.

\begin{figure} %
\centering
\includegraphics[width=0.5\textwidth]{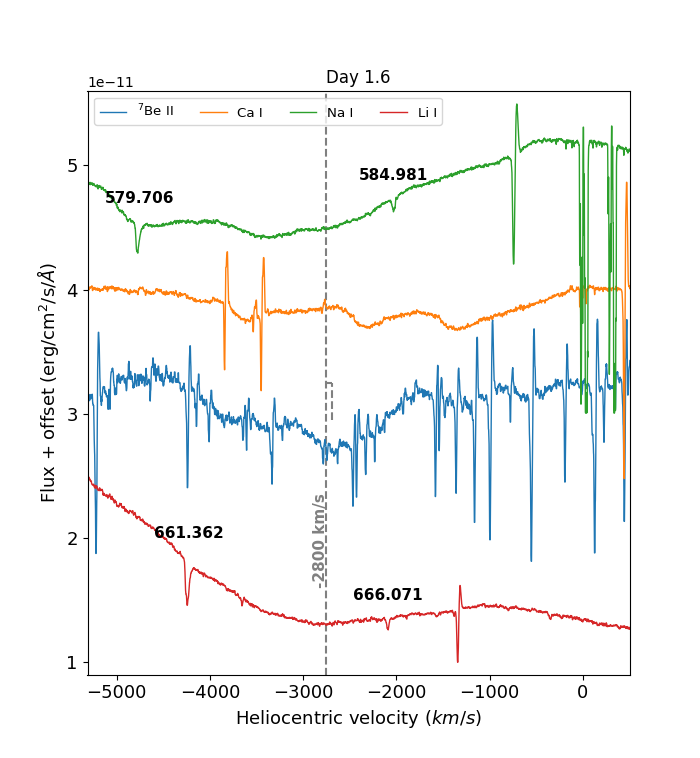}
\caption{ Spectrum of RS Oph   in the regions of \liviii\ 670.8 nm, the D2 line and \nai\ 589.0 nm and \cai\ 422.6 nm.   In the portion around  \liviii\ 670.8 nm the narrow absorptions are due to the DIB 666.071 nm and 661.362 nm while in the \nai\ region there are the DIBs at 584.981 and 579.706 nm. The P-Cygni lines in the other spectra are due to metal lines arising in the red giant wind.   The    \hei\   587.6
nm and   \hei\ 667.8 nm  are present in the \nai\  and  \lii\  spectral regions, respectively. }\label{fig11}
\end{figure}

\subsection{Ejected mass}

A direct method to estimate the mass of  the ejecta consists in deriving it
by using  the emission intensity   of a recombination line
together with a
good estimate of  the  electron  density in the shell during the nebular phase.
A suitable line is  ${H_{\beta}}$ and its    luminosity  at these epochs can be expressed as 

\begin{eqnarray}\label{eq:2}
  L_{H_{\beta}}
   =   3.03 \cdot 10^{-14} \cdot h{\nu} \cdot N_e \cdot N_p \cdot V_{ej} \cdot f  &&[erg  \cdot s^{-1}] 
\end{eqnarray}

 where $3.03 \cdot 10^{-14}$ is the $H_{\beta}$ effective recombination coefficient \citep{Osterbrock2006},    $h{\nu}$ is the photon energy, $N_e$ and  $N_p$ are the electron and proton number densities respectively,  while $ V_{ej}$  and  f are the volume and  filling factor, respectively. The ejected mass can be written as

\begin{eqnarray}\label{eq:3}
  M_{ej}
  & \approx  &  \mu \cdot m(H) \cdot N_p \cdot V_{ej} \cdot f  
\end{eqnarray}

where m(H)= $1.67 \cdot 10^{-24}$ grams  and $\mu \approx 1.4$ is the mean molecular weight.

 By combining the two above equations, which has the advantage of cancelling both volume and filling factor, we obtain:

\begin{eqnarray}\label{eq:4}
  M_{ej}
  & = & \frac{ 18.5 \cdot   L_{H_{\beta}}}{ N_e}     
\end{eqnarray}

\begin{figure*} %
\centering
\includegraphics[width=1.05\textwidth]{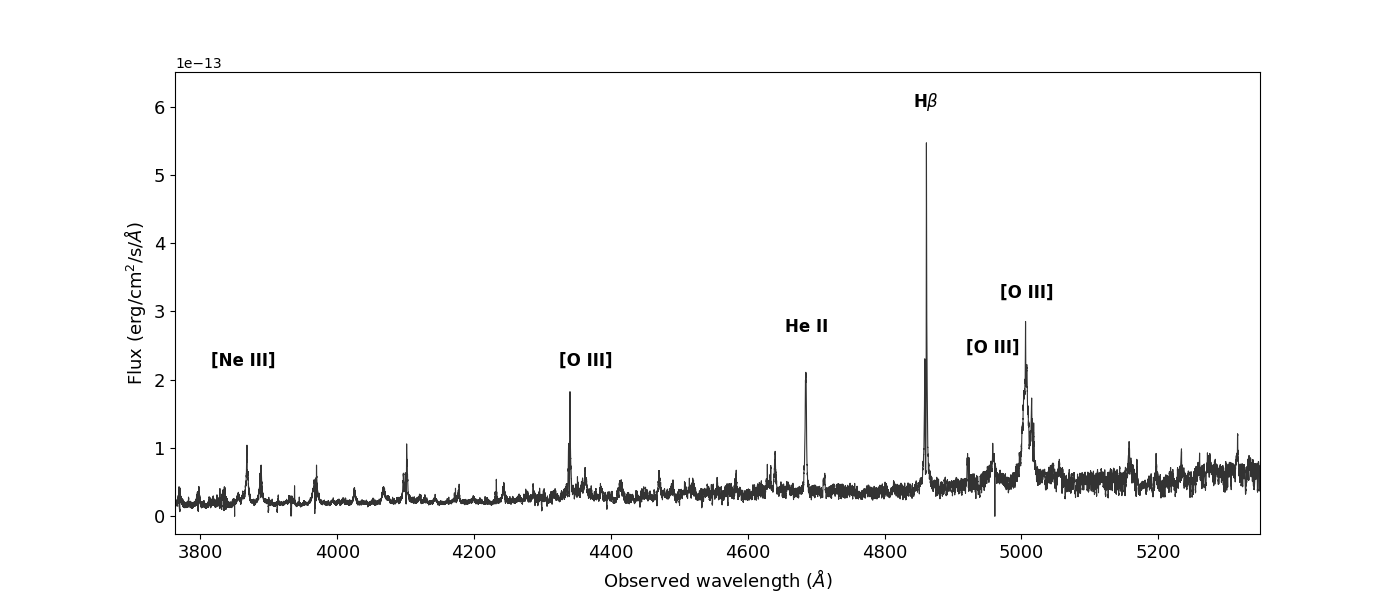}
\caption{The spectrum of RS Oph obtained on 29 March 2022 in the range (380-530) nm. Nebular emission lines are detected on top of a  continuum which shows signatures of  the red giant atmosphere. The position of the most common ones and the lines used for the estimate of the ejected mass are marked.}\label{fig12a}
\end{figure*}

We  obtained several spectra of the nova at the start of the  nebular phase. In  Fig. \ref{fig12a}, we show the spectrum of 29 March 2022 obtained with UVES. We used a large slit of 3 arcsec to collect  all the light ($\sim 97\%$) required  to obtain an accurate spectro-photometric flux calibration. 
The observed   flux of $H_{\beta}$ in the spectrum corrected for  reddening  $E(B-V) = 0.73$ \citep{cassatella1985ESASP.236..281C,snijders1987Ap&SS.130..243S} is of $H_{\beta}
     = 11.5 \cdot  10^{-12}$   $erg \cdot  s^{-1} \cdot cm^{-2}$.   At the distance of 2400 pc it becomes $L_{H_{\beta}}    =  7.95 \cdot  10^{33}$  $erg \cdot s^{-1}$. 
     and therefore we obtain:

\begin{eqnarray}\label{eq:4}
  M_{ej}
  & = &  \frac{74.9}{N_e}  ~~~~~[M_{\odot}]    
\end{eqnarray}

The electron density in the optical range  is generally  derived from the
ratio of two closely spaced   emission lines ,  e.g.  [\oii]
 $\frac{372.9}{372.6}$   and [\sii]  $\frac{671.6}{673.1}$.  
Unfortunately, these emission lines are not present in the  March 
spectra.  However, the electron
density  can be derived also from the  ratio of the   [\oiii] lines   $\frac{495.9+500.7}{436.3}$.
This ratio is generally a function of both T$_e$ and N$_e$, but for N$_e$
greater than $10^4$ $cm^{-3}$  the ratio varies strongly with N$_e$  because
the upper level $^1D_2$  of the  495.9 nm and 500.7 nm  lines begins to get
collisionally de-excited.
Under this condition  of high density, the T-Ne
diagnostic  equation can be represented as \citep{Osterbrock2006}:

\begin{eqnarray}\label{eq:5}
  \frac{J_{495.9} + J_{500.7}}{J_{436.3}}
   & = &  \frac{ 7.90 \cdot e^{[(3.29\cdot   10^4)/T_{e}]}} { 1 + 4.5 \cdot 10^{-4} ( N_e/T_{e}^{1/2})} .    
\end{eqnarray}

For    $T_e = 10^4$ K, a common value  in  photoionized nebulae,  and inserting the values reported in Tab. \ref{tab4}, the ratio  is 16.9 and 
    $ N_e \approx 2.57 \cdot 10^6 $.

    This value is rather stable, assuming $T_e = 10^4$ K, the ejected mass based on $H_{\beta}$  is of M$_{ej} \approx  2.9 \cdot 10^{-5}  M_\odot$.  and a  $T_e =8\cdot  10^3$ K would give  $ N_e \approx 6.5 \cdot 10^6 $, while a $T_e =1.2 \cdot  10^4$ K would give  $ N_e \approx 1.4 \cdot 10^6 $ corresponding in an uncertainty in the mass of  ejecta of about a factor two.
   
\begin{table}
\caption{ Nebular Lines used for the estimation of the nova ejecta as measured in the spectrum of 29 March 2022. F$_c$ is the flux of the continuum adjacent to the emission lines and F is the total flux of the line above the continuum.  }
\label{tab4}
\scriptsize
\begin{tabular}{lrrrr}
\hline
\hline
\multicolumn{1}{c}{{ident}} &
\multicolumn{1}{c}{{$\lambda_{obs}$}} & 
\multicolumn{1}{c}{{FWHM}} & 
\multicolumn{1}{c}{{F$_c$}} &
\multicolumn{1}{c}{{F}}
 \\
 \multicolumn{1}{c}{ }&
\multicolumn{1}{c}{{nm}} & 
\multicolumn{1}{c}{{\AA}} &
\multicolumn{1}{c}{{erg cm$^{-2} s^{-1}$}} &
\multicolumn{1}{c}{{erg cm$^{-2} s^{-1}$}}  
 \\
\hline
\hline
H$_\beta$  486.135      & $\approx$ 485.985           & 2.81   & 4.9 $\cdot 10^{-13}$& 1.15 $\cdot 10^{-11}$\\
\heii\ 468.6571 & 468.522 &2.73 &3.8 $\cdot 10^{-13}$ &5.69 $\cdot 10^{-12}$\\
\oiii\ 	436.3209 &436.2199	&3.22   &  4.2 $\cdot 10^{-13}$       & 1.24 $\cdot 10^{-12}$    \\
\oiii\ 	495.8911 &495.8654	&9.23  &  3.9 $\cdot 10^{-13}$        & 5.50 $\cdot 10^{-12}$    \\
\oiii\ 	500.6843 &500.6074	 &  9.56         & 4.4 $\cdot 10^{-13}$& 1.55 $\cdot 10^{-11}$    \\
\hline
\hline
\end{tabular}
\end{table}

  However, H$_{\beta}$ is found in emission also during  quiescence and it is also  difficult to separate the contribution of the outburst. The value derived above is more strictly  an upper limit to the value of the mass of the ejecta. We have therefore used  the \heii\ 468.6 nm which has a  certain nebular origin. The equations for \heii\ become:

\begin{eqnarray}\label{eq:2}
  L_{\ion{He}{II}} 
  =   1.58 \cdot 10^{-24} \cdot N_e \cdot N_{HeII} \cdot V_{ej} \cdot f  &&[erg  \cdot s^{-1}] 
\end{eqnarray}
and

\begin{eqnarray}\label{eq:3}
  M_{HeII}
  & \approx  &  \mu  \cdot m(H) \cdot N_{HeII} \cdot V_{ej} \cdot f  
\end{eqnarray}

and we obtain

\begin{eqnarray}\label{eq:4}
  M_{HeII}
  & = & \frac{ 1.5 \cdot   L_{\ion{He}{II}}}{N_e}     
\end{eqnarray}

The observed flux of \heii\ 468.6 nm as measured in the spectrum of 29 March 2022  is
L$_{\heii}$ = 3.0 $\cdot  10^{33}$  erg $\cdot s^{-1}$ for a  distance of 2.4 Kpc.
For $ N_e \approx 2.57 \cdot 10^6 $ and considering negligible   the presence of \hei\  in the same layers    the helium mass  is of M$_{HeII} = 1.73 \cdot 10^{27}$ grams or $ 8.7 \cdot 10^{-7} M_\odot$ and using a number ratio of  H/He = 12.6   we obtain an ejected mass     of M$_{ej} \approx   1.1 \cdot 10^{-5} M_\odot$.
  
  \citet{ruchi2022arXiv220710473P} estimated of the mass of the ejecta in the 2021 outburst  with a Cloudy model for the ejecta with hydrogen density, volume and filling factor derived by the best model. These authors  estimate the mass of M$_{ej} \approx   3-4 \cdot 10^{-6} M_\odot$  but for  a distance of  1.68 kpc. By adopting the GAIA distance of  ~ 2.4 kpc we are using here  the  ejecta would be  a factor 1.4 higher and about a factor two lower of what derived here. 

For the  outburst of 2006 \citet{das2006ApJ...653L.141D,das2015NewA...39...19D} 
derived an ejected mass of M$_{ej} \approx   3-5 \cdot 10^{-6} M_\odot$.  Modeling the x-ray emission from Chandra observations
\citet{orlando2009A&A...493.1049O} derived  M$_{ej} \approx   10^{-6} M_\odot$. \citet{eyres2009MNRAS.395.1533E} derived M$_{ej} \ge   4 \cdot 10^{-7} M_\odot$ and \citet{vaytet2011ApJ...740....5V}  M$_{ej} \approx   2-5 \cdot 10^{-7} M_\odot $. For the  outburst of 1985 \citet{anupama1989JApA...10..237A} derived a M$_{ej} \approx   2.3-3.7 \cdot 10^{-6} M_\odot $.

 An ejected  mass of $\approx 1.1 \cdot  10^{-5}$ M$_\odot$    is a relatively high value when compared to the typical ejected material of a RN, and  similar to the mass of  CNe ejecta,   \citep{dellavalle2020A&ARv..28....3D}. Historical observations of RNe in M31  showed  that RNe might be outliers of the Maximum Magnitude vs Rate of Decline (MMRD) relationship \citep{arp1956AJ.....61...15A,rosino1973A&AS....9..347R}. 
An effective  test to verify  the mass of the RS Oph ejecta  can be performed by studying the position of  RS Oph in the MMRD plane. After correction for extinction RS Oph reached V = 2.7 at  peak  which  at  the distance of d = 2.40 $\pm$ 0.16 Kpc  corresponds to M$_V \approx$  -9. From the lightcurve shown in Fig. \ref{fig1}  we get $T_2$ = 4.2 days and a decline of $ \delta V$ = 0.48 mag/day. These data show that RS Oph matches  well the MMRD  of CNe in agreement with our measurement of the RS Oph ejected mass \citep{dellavalle1991A&A...252L...9D}.

We note interesting  consequences of  such  high ejecta.  RS Oph is likely to have expelled more material than it has been able to accumulate   since the 2006 outburst. Should this  be confirmed by the study of previous and future outbursts, it would imply that the WD  is eroding during its duty-cycle, thus preventing it  from  reaching Chandrasekhar  mass and exploding as a type-Ia SN.


\subsection{Lithium}

 In Fig. \ref{fig11}  the  spectrum of RS Oph  1.6 day after explosion  is shown in the region of \liviii\ 670.8 nm,  the  D2 line of \nai\ 589.0 nm and \cai\ 422.6 nm. No absorption is detected in correspondence of these neutral species  in this and also all other epochs. The narrow absorptions  seen in the spectra are due to DIBs while the P-Cygni profile are due to lines originated in the red-giant wind.  In particular, there is no  evidence of the resonance \liviii\ doublet at 670.8 nm in the spectra of RS Oph  as in   most  classical novae. So far,  \liviii\ has been detected only in  V1369 Cen \citep{Izzo2015}.
 The  detection in  V1369 Cen  was determined  on days 7 and 13 when  only a small amount of \bevii\  decayed into \livii,  which  implies that the TNR started much  earlier than the explosion \citep{Izzo2015}. Li has been also detected  as a trace element  in a very early epoch of    V906 Car  \citep{molaro2020MNRAS.492.4975M} and in V5668 Sgr \citep{Izzo2019}. The equivalent width of the $^7$Li line in V5668 Sgr declines with time as the ionization of the ejecta increases and disappears by day 42 after explosion  \citep{wagner2018AAS...23135810W}.
A claim for  \liviii\ 670.7 nm detection  in V382 Vel was made by \cite{DellaValle2002},  but  \cite{Shore2003AJ....125.1507S} suggested that it  could be neutral nitrogen. Neutral $^7$Li remains absent from nova outburst spectra also  when  observations extend  to a time scale longer than the \bevii\ decay. This is the reason why novae have not been recognized as Li producers for decades after the theoretical suggestion. However, this can  be  explained if  \bevii\ decays through the   capture of an internal K-electron  to  end as  ionized lithium  with   transitions only  in the X-ray domain \citep{Molaro2016}.
The    \liviii\ line was  discovered in quiescent spectra  of RS Oph and T CrB 
 by \citet{wallerstein2008PASP..120..492W}. The line shows  the orbital motion and therefore it is formed  in  the atmosphere of the red giant \citep{brandi2009A&A...497..815B}. The abundance is of A(Li) = 1.1, but if we consider  the effective temperature and luminosity
of this star, \livii\  should have been completely destroyed \citep[see e.g.][and references therein]{Lambert80,Charbonnel20,Magrini21}.
This fact was already noted by \citet{wallerstein2008PASP..120..492W} both
for RS Oph and for T CrB. Although Li-rich giants exist they are
rare \citep[see e.g.][and references therein]{Kumar_2011,Deepak19,Martell21}.
There is an unsettled debate on whether the \livii\ in Li-rich giants is intrinsic (i.e. produced in the star itself) or extrinsic (i.e. produced elsewhere and then accreted onto the star). \citet{Kumar20} argued that all stars must undergo a \livii\ production phase between the tip of the Red Giant Branch and the Red Clump. The time spent in this phase is short and only large surveys allow to detect these stars.
From the observational point of view we remark that it is surprising that
two RNe would  show measurable \livii\ in the atmosphere of the cool red giant companion. It is tempting to conjecture that at each outburst a part of the
freshly produced \livii\ in the TNR is accreted by the companion star building up
and compensating for the destruction due to the convection a measurable
amount of \livii.

\begin{figure}
\centering
\includegraphics[width=0.5\textwidth]{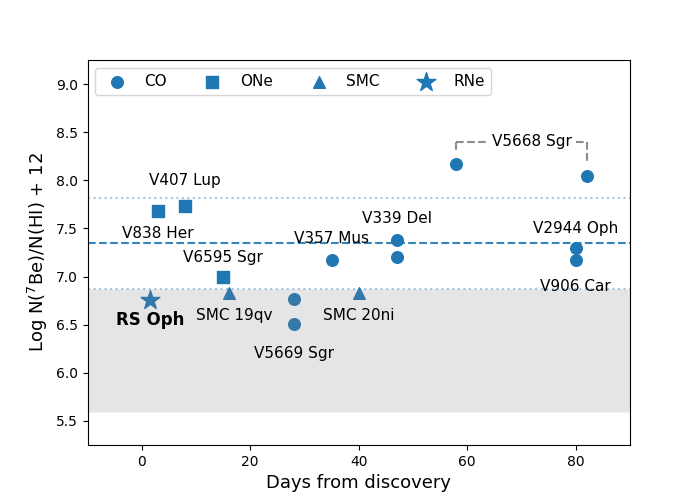}
\caption{  \bevii\  abundance in RS Oph  and in all  CNe where it was found. The plot shows in  ordinate the  A(\bevii) = $log N(^7Be)/N(HI) +12  $, with A(\bevii) = 6.76 for RS Oph. Fast novae are marked with squares, slow novae with circles while MC novae with triangles.  The dashed line mark the average value, estimated using only MW novae.  To note that the meteoritic    abundance is A(\livii) = 3.3  \citep{Lodders2009}.
}\label{fig13}
\end{figure}

\section{Discussion}

The possible presence of \bevii\  in the outburst
spectra of  RS Oph adds a member of the class of Recurrent Novae  to the small sample of objects where \bevii\  has been detected.  The A(\bevii) \footnote{A(\bevii) = $log N(^7Be)/N(HI) +12  $} yields in RS Oph and   the ones measured in the  classical novae are compiled
 in Tab. \ref{tab5} and are shown in Fig. \ref{fig13}.  CNe   show  a range of different \bevii\  yields scattered   by one order of magnitude. The   scatter  exceeds the admittedly large observational errors and is likely real. The light mass of the ejecta and the high terminal velocity is typical of ONe novae but  \citet{mikolajewska2017ApJ...847...99M} classified  RS Oph as a CO  nova. However,  no difference in the yields between CO and ONe novae is found at variance with theoretical predictions. The mean value is of A(\bevii) = 7.34 $\pm 0.47 $ while in RS Oph the abundance is of 6.78 overlapping with the lowest abundances derived in CNe. 

Assuming the Li meteoritic abundance  as the best proxy for the present Li value in the Galaxy, the current  mass of  \livii\  can  be  estimated  to be  about  1000 M$_\odot$ \citep{starrfield2020ApJ...895...70S}.   About 10-25 \%\  of this  has been  made in the  primordial nucleosynthesis and another  10\%   by the slow spallation processes taking place in the interstellar medium along the whole Galactic life. AGB stars could contribute at most  by a  few percent  \citep{Romano2001}. The astronomical source for the remaining 70\%  remains to be identified. In fact, the real fraction is even higher  considering that about 20-30 percent of Li is  burned  through  recycling within stars.

 Detection of significant overabundances of \bevii\ in Classical novae, with yields that are about 4 orders
of magnitude over  the meteoritic \livii\ value \citep{lodders2019arXiv191200844L},  makes these systems the more  plausible candidates  for making   the missing       Galactic \livii\ fraction.
 Detailed models of the chemical evolution of
the Milky Way   showed that novae account well for
the observed increase of Li abundance with metallicity in the
 thin disk, and also for the relative flatness observed
in the thick disk 
\citep{Cescutti2019}. In fact, the thick disk evolves on a timescale
which is shorter than the typical timescale for the production
of substantial \livii\  by novae. \cite{Cescutti2019} left the
nova yields as a free parameter and found that assuming a  nova undergoes explosion every $ 10^{4}$ yr, as suggested by \citet{ford1978ApJ...219..595F}, in order to match
the \livii\  growth   and the present abundance,
a \livii\  production of  3$ \cdot 10^{-9} M_\odot$   per nova event is required, which is about what  observed.

 Detection of \bevii\ in recurrent novae as RS Oph provides further support to the suggestion that Novae are the main source of lithium.  
   The   ejected   mass   in the RS Oph 2021 outburst 
 is  probably $ \ge 6 \cdot 10^{-6}$  M$_{\odot}$. With a measured yield in mass of $X(\mbox{\bevii})/X(\mbox{H})$  
= 4.0 $\cdot 10^{-5}$, the amount of \livii\   created in the RS Oph 2021  event  is
of $\ge$  2.4 $\cdot $ 10$^{-10}$ M$_{\odot}$. 

 Typical RN could have  ejecta  by one or two order of magnitude
smaller than RS OPH and therefore  could synthesize   something between  10$^{-1}$ to 10$^{-3}$  of the average   \livii\   of  CNe  per event  but  occur 10$^3$ times more frequently
 \citep{ford1978ApJ...219..595F}.   The   fraction of RNe  is difficult to know from observations   due to the limited time  of  astronomical observations and cannot be excluded that some of the nova observed for \bevii\ is in fact a RN. In fact, all binaries with a WD  are likely recurrent novae with a period which is primarily determined by the mass of the WD and by the mass transfer from the  companion. \citet{dellavalle1996ApJ...473..240D} have estimated for M31 and LMC a fraction RN/CN of 10\% and 30\%.  In the Milky Way a ratio  of  12- 35\%  has been found by \citet{pagnotta2014ApJ...788..164P} and of 30\% by \citet{dellavalle2020A&ARv..28....3D}.
 Thus, RNe could  be  one third of the CNe and  could   concur in    making   of  the \livii\ we observe today in the Milky Way.

\begin{table*}
\caption{   
A(\bevii)  
abundances    for the  CN and RS Oph. 
$N($\bevii$)/N(H)_c$ are the values   corrected for the
 \bevii\ decay with a mean life of 76.8 days.}
\label{tab5}
\begin{center}
\begin{tabular}{llrrrrll}
\hline
\hline
\multicolumn{1}{c}{Nova} &
\multicolumn{1}{c}{type} &
\multicolumn{1}{c}{d} &
\multicolumn{1}{c}{comp} &
\multicolumn{1}{c}{ A(\bevii) } &
\multicolumn{1}{c}{ A(\bevii)$_c$ } &
\multicolumn{1}{c}{Ref}& \\
\multicolumn{1}{c}{} &
\multicolumn{1}{c}{} &
\multicolumn{1}{c}{} &
\multicolumn{1}{c}{\kms} &
\multicolumn{1}{c}{  } &
\multicolumn{1}{c}{} &
\multicolumn{1}{c}{}\\
\hline
V339 Del & CO & 47 & -1103& 6.92 &7.20  & \cite{Tajitsu2015},\cite{Tajitsu2016}  \\
V339 Del & CO &47  &-1268 &  7.11   & 7.38   & \cite{Tajitsu2015},\cite{Tajitsu2016}  \\
V5668 Sgr & CO &  58 & -1175 &  7.84 & 8.17   & \cite{Molaro2016}    \\
 V5668 Sgr& CO & 82 &-1500 &7.58 &8.04  & \cite{Molaro2016}   \\
 V2944 Oph  & CO &  80 &  -645 & 6.72  & 7.18  & \cite{Tajitsu2016} \\
V407 Lup  & ONe&  8  &-2030 & 7.69     & 7.73  & \cite{Izzo2018}  \\
V838 Her & ONe?&  3  & -2500 &   7.66  & 7.68 &\cite{Selvelli2018}  \footnotemark[1] \\
V612 Sct & ?&    &  & &  - &  \cite{molaro2020MNRAS.492.4975M} \\
V357 Mus &CO? &  35  & $\approx$  -1000 & 6.96& 7.18&   \cite{molaro2020MNRAS.492.4975M} \\
FM Cir & CO?&    & &   &: &  \cite{molaro2020MNRAS.492.4975M} \\
V906 Car & CO?&  80  & $\approx$ -600 & 6.86 & 7.30& \cite{molaro2020MNRAS.492.4975M}   \\
V5669 Sgr & CO & 28& $\approx$ -1000 &  6.34& 6.51 & \cite{Arai2021ApJ...916...44A}  \\
V5669 Sgr & CO & 28& $\approx$ -2000 & 6.61 & 6.77& \cite{Arai2021ApJ...916...44A} \\
V6595 Sgr & ONe & 15  &-2700  & 6.87  & 6.99  &  \cite{molaro2021arXiv211101469M}\\
\hline
V1369 Cen & CO & 7& -550 & 5.00 & 5.04 & \cite{Izzo2015} \footnotemark[2] \\
V1369 Cen & CO & 13& -560 & 5.30 & 5.38 & \cite{Izzo2015} \footnotemark[2]\\
V1369 Cen$^*$ & CO & 7& -550  &4.70 & 4.78 & \cite{Izzo2015}\footnotemark[3]\\
V1369 Cen$^*$ & CO & 13& -560  &4.78 & 4.85 & \cite{Izzo2015} \footnotemark[3]\\
\hline
SMC ASASSN-19qv & CO & 16 & -2400 & 6.62 &6.71& \cite{Izzo2021MNRAS.tmp.3442I}\\
SMC ASASSN-20ni & CO? & 40 & $\approx$ -520 & 6.41 &6.73&\cite{Izzo2021MNRAS.tmp.3442I}\\
\hline
RS Oph &NR &1.6&-2800&6.78&6.78&this paper\\
\hline
\end{tabular}
\footnotetext{1. The measurement from \mgii\ absorption is reported.}
\footnotetext{2. The \livii\ abundance is measured  from   \liviii\ 670.8 nm  and ionization is estimated with  \nai}
\footnotetext{3. The \livii\ abundance is measured  from   \liviii\ 670.8 nm  and ionization is estimated with      \ki}
\end{center}

\end{table*}

\section{Conclusions}\label{sec13}

\begin{itemize}

\item{By means of  the high resolution UVES spectrograph at the Kueyen-UT2 telescope of the ESO-VLT, Paranal, HARPS-N at the TNG and FIES at the NOT, both in La Palma, Spain, we  monitored the 2021 outburst of the recurrent nova RS Oph since day 1.6 from the explosion.  
Analysis at high resolution in the far blue spectral region enabled  detection of   the possible presence of \bevii\ at 313.0 nm freshly made in the thermonuclear runaway reactions showing that  CNe and  RNe behave similarly.}

\item{ The  \bevii\ yields can be estimated from the analysis of the first spectrum where the \caii\ K line is present and is not affected by the H emission. 
From the fourth day after the explosion the metal lines  are no longer visible in absorption, probably due to the low mass  of the ejecta.
The yields measured are of    N(\bevii)/N(H) 
= 5.7 $\cdot 10^{-6}$   which are     close to the  lowest values  measured 
in classical novae.}

\item{
 By means of spectra taken in the nebular phase  we estimate a  mass ejecta of M$_{ej} \approx   1.1 \cdot 10^{-5}  M_\odot$  providing an  amount of $\approx$  4.4 $\cdot $ 10$^{-10}$ M$_{\odot}$ of \livii\   created in the 2021 event. }

 \item{  
   RNe of the kind of RS Oph synthesize  around  the same   \livii\  per event than  classical novae,   but  occur 10$^3$ times more frequently.  Recurrent Novae could be one third of classical novae and, therefore,  they  could have concurred significantly, depending on their ejected mass, to the  making   of   \livii\ we observe today in the Milky Way. The detection of \bevii\  in RS Oph provides further support to the recent claim that novae are the main source of \livii.
 }
 \end{itemize}

\section*{Acknowledgments}

The observations have been taken under a Target opportunity Program 105.20B6.001, 105.20B6.002 P.I. Paolo Molaro. 
 The ESO staff is warmly acknowledged for the execution of these observations  during the pandemic  lockdown.  Some  observations are from the ToO A3TAC20, P.I. L. Izzo and were  made with the Italian Telescopio Nazionale Galileo (TNG) operated on the island of La Palma by the Fundaciòn Galileo Galilei of  INAF (Istituto Nazionale di Astrofisica) at the Spanish Observatorio del Roque de los Muchachos of the Instituto de Astrofisica de Canarias. Nando Patat is warmly thanked for discussions on the red giant wind and for  making available his spectra of RS Oph in quiescence.  LI was supported by two grants from VILLUM FONDEN (project number 16599 and 25501). Based on observations made with the Nordic Optical Telescope, owned in collaboration by the University of Turku and Aarhus University, and operated jointly by Aarhus University, the University of Turku and the University of Oslo, representing Denmark, Finland and Norway, the University of Iceland and Stockholm University at the Observatorio del Roque de los Muchachos, La Palma, Spain, of the Instituto de Astrofisica de Canarias. MH acknowledges funding support from the MICIN/AEI grant PID2019-108709GB-I00.

\section{data availability}
 Based on  data from the UVES spectrograph at the Unit 2 of the VLT at the Paranal Observatory, ESO, Chile.   ESO data are world-wide available and can be requested after the proprietary period of one year by the astronomical community through the link http://archive.eso.org/cms/eso-data.html. They will be shared earlier on  reasonable request with  the corresponding author.

\bibliographystyle{mnras}
\bibliography{rsoph_mnras.bib}
\end{document}